\documentclass[conference,english,HTML,table]{IEEEtran}
\IEEEoverridecommandlockouts

\usepackage[utf8]{inputenc}
\usepackage[T1]{fontenc}
\usepackage{cite}
\usepackage{amsmath,amssymb,amsfonts}
\usepackage{algorithmic}
\usepackage{array}
\usepackage{babel}
\usepackage{booktabs}
\usepackage{CJKutf8}%
\usepackage[abbreviations]{foreign}
\usepackage{graphicx}
\usepackage{ifthen}
\usepackage{multirow}
\usepackage{pifont}
\usepackage{textcomp}
\usepackage{tcolorbox}
\usepackage{tikz}
\usepackage{url}
\usepackage{xspace}
\GetGinDriver%
\usepackage[\GinDriver,hidelinks]{hyperref}%
\usepackage[translate=true]{glossaries}%
\usepackage{glossaries-extra}
\usepackage{cleveref}%

\def\BibTeX{{\rm B\kern-.05em{\sc i\kern-.025em b}\kern-.08em
    T\kern-.1667em\lower.7ex\hbox{E}\kern-.125emX}}
\graphicspath{{./figures/}}
\newabbreviation{afl}{AFL}{american fuzzy lop}
\newabbreviation{api}{API}{application programming interface}
\newabbreviation{ccsc}{CCSC}{common component security constraint}
\newabbreviation{ci}{CI}{continuous integration}
\newabbreviation{cia}{CIA}{confidentiality, integrity, availability}
\newabbreviation{cicd}{CI/CD}{continuous integration / continuous delivery}
\newabbreviation{cli}{CLI}{command line interface}
\newabbreviation{cr}{CR}{component requirement}
\newabbreviation{crt}{CRT}{communication robustness testing}
\newabbreviation{csa}{CSA}{component security assurance}
\newabbreviation{cvss}{CVSS}{common vulnerability scoring system}
\newabbreviation{cwe}{CWE}{common weakness enumeration}
\newabbreviation{dos}{DoS}{denial of service}
\newabbreviation[longplural={devices under test}]{dut}{DUT}{device under test}
\newabbreviation{edsa}{EDSA}{embedded device security assurance}
\newabbreviation{ert}{ERT}{embedded robustness testing}
\newabbreviation{fr}{FR}{foundational requirement}
\newabbreviation{fsa}{FSA}{functional security assessment}
\newabbreviation{gvm}{GVM}{Greenbone Vulnerability Management}
\newabbreviation{hmi}{HMI}{human-machine interface}
\newabbreviation{iacs}{IACS}{industrial automation and control system}
\newabbreviation{ide}{IDE}{integrated development environment}
\newabbreviation{iiot}{IIoT}{industrial IoT}
\newabbreviation{iot}{IoT}{Internet of things}
\newabbreviation{isci}{ISCI}{ISA Security Compliance Institute}
\newabbreviation{it}{IT}{information technology}
\newabbreviation{nvd}{NVD}{national vulnerability database}
\newabbreviation{os}{OS}{operating system}
\newabbreviation{oss}{OSS}{open-source security}
\newabbreviation{ot}{OT}{operational technology}
\newabbreviation{re}{RE}{requirement enhancement}
\newabbreviation{sda}{SDA}{security development artifacts}
\newabbreviation{sdk}{SDK}{software development kit}
\newabbreviation{sdla}{SDLA}{security development lifecycle assurance}
\newabbreviation{sl}{SL}{security level}
\newabbreviation{sl-c}{SL-C}{capability security level}
\newabbreviation{srt}{SRT}{system robustness testing}
\newabbreviation{sut}{SUT}{system under test}
\newabbreviation{vit}{VIT}{vulnerability identification testing}

\makenoidxglossaries
\definecolor{LegendBack}{HTML}{DEEBF7}
\definecolor{LegendBorder}{HTML}{3182BD}
\newcommand*{\fillcirc}[1]{\raisebox{-1pt}{\begin{tikzpicture}[scale=0.15]%
    \filldraw[color=black,fill=#1] (0,0) circle (0.6);
    \end{tikzpicture}}}
\newcommand*{\filltextcirc}[1]{\raisebox{-1pt}{\begin{tikzpicture}[scale=0.15]%
    \filldraw[color=black,fill=#1] (0,0) circle (0.8);
    \end{tikzpicture}}}
\newcommand*{\fillsquare}[1]{\raisebox{-1pt}{\begin{tikzpicture}[scale=0.15]%
    \filldraw[color=black,fill=#1] (0,0) rectangle (1.6,1.6);
    \end{tikzpicture}}}

\definecolor{PuOr3Orange}{HTML}{F1A340}
\definecolor{PuOr3White}{HTML}{F7F7F7}
\definecolor{PuOr3Violet}{HTML}{998EC3}
\definecolor{GreysBlack}{HTML}{636363}
\definecolor{GreysGray}{HTML}{BDBDBD}
\definecolor{GreysWhite}{HTML}{F0F0F0}

\newcommand{\covered}{\fillcirc{PuOr3Violet}}
\newcommand{\covpartially}{\fillcirc{PuOr3White}}
\newcommand{\uncovered}{\fillcirc{PuOr3Orange}}
\newcommand{\textcovered}{\filltextcirc{PuOr3Violet}}
\newcommand{\textcovpartially}{\filltextcirc{PuOr3White}}
\newcommand{\textuncovered}{\filltextcirc{PuOr3Orange}}
\newcommand{\blackbox}{\fillsquare{black}}
\newcommand{\graybox}{\fillsquare{GreysGray}}
\newcommand{\whitebox}{\fillsquare{white}}
\newcommand{\nracheron}{\ding{202}}
\newcommand{\nrachilles}{\ding{203}}
\newcommand{\nrbestorm}{\ding{204}}
\newcommand{\nrdefensics}{\ding{205}}
\newcommand{\nrnessus}{\ding{206}}
\newcommand{\nrraven}{\ding{207}}
\newcommand{\nrvhunter}{\ding{208}}
\newcommand{\nrafl}{\ding{192}}
\newcommand{\nrburpsuite}{\ding{193}}
\newcommand{\nrjmeter}{\ding{194}}
\newcommand{\nrnmap}{\ding{195}}
\newcommand{\nropenvas}{\ding{196}}
\newcommand{\nrscapy}{\ding{197}}
\newcommand{\nrselenium}{\ding{198}}
\newcommand{\nrsymcc}{\ding{199}}

\newcommand\copyrighttext{\footnotesize \textcopyright 2023 IEEE.
Personal use of this material is permitted.
Permission from IEEE must be obtained for all other uses, in any current or future media, including reprinting/republishing this material for advertising or promotional purposes, creating new collective works, for resale or redistribution to servers or lists, or reuse of any copyrighted component of this work in other works.
Presented in the \href{https://2023.ieee-etfa.org/}{2023 IEEE 28th International Conference on Emerging Technologies and Factory Automation (ETFA)}.
The final version of this paper is available under DOI: \href{https://doi.org/10.1109/ETFA54631.2023.10275637}{10.1109/ETFA54631.2023.10275637}%
}
\newcommand\copyrightnotice{\begin{tikzpicture}[remember picture,overlay]
\node[anchor=south,yshift=10pt,fill=yellow!20] at (current page.south) {\fbox{\parbox{\dimexpr\textwidth-\fboxsep-\fboxrule\relax}{\copyrighttext}}};
\end{tikzpicture}}

\newboolean{tv2}
\setboolean{tv2}{false}

\date{}
\title{Qualitative Analysis for Validating IEC 62443-4-2 Requirements in DevSecOps}
\author{
Christian Göttel,
Maëlle Kabir-Querrec,
David Kozhaya,
Thanikesavan Sivanthi,
Ognjen Vuković,
\\
\IEEEauthorblockA{
ABB Schweiz AG, Corporate Research Center, Baden-Dättwil, Switzerland, \texttt{first.last@ch.abb.com}
}
}

\begin{document}

\maketitle
\copyrightnotice

\begin{abstract}
Validation of conformance to cybersecurity standards for industrial automation and control systems is an expensive and time consuming process which can delay the time to market. It is therefore crucial to introduce conformance validation stages into the continuous integration/continuous delivery pipeline of products. However, designing such conformance validation in an automated fashion is a highly non-trivial task that requires expert knowledge and depends upon available security tools, ease of integration into the DevOps pipeline, as well as support for IT and OT interfaces and protocols.

This paper addresses the aforementioned problem focusing on the automated validation of ISA/IEC 62443-4-2 standard component requirements. We present an extensive qualitative analysis of the standard requirements and the current tooling landscape to perform validation. Our analysis demonstrates the coverage established by the currently available tools and sheds light on current gaps to achieve full automation and coverage. Furthermore, we showcase for every component requirement where in the CI/CD pipeline stage it is recommended to test it and the tools to do so.

\end{abstract}

\section{Introduction}
\label{sec:intro}

\Gls{iacs} have to perform their mission-critical tasks in a safe and secure manner with high availability. The prevailing trend in \gls{iacs} is the convergence of \gls{ot} and \gls{it} networks and systems. This increases the overall attack
surface of \gls{iacs}, which are not only exposed to acts by curious insiders or unintended errors of genuine users, but also to malicious cyber attacks from the outside.
Cyber attacks on such systems may have as consequences loss of data or interruption of operations, and may as well span beyond the targeted organizations by compromising the safety, health and environment of entire regions or nations.
As cyber threats pose a growing risk to \gls{iacs}, there is an increasing demand for cybersecurity to mitigate cyber attacks. The industry has developed standards such as ISA/IEC 62443~\cite{iec:62443-1-1} to help product vendors, integrators and operators secure \gls{iacs}. 

\begin{figure}[!t]
    \centering
    \includegraphics[width=.67\linewidth]{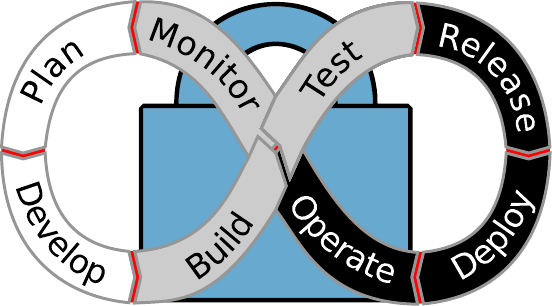}
    \caption{DevSecOps with separation of the DevOps lifecycle into white-box (\protect\whitebox), gray-box (\protect\graybox), and black-box (\protect\blackbox) testing methods.\label{fig:overview}}
\end{figure}

DevOps, which automates and integrates the work of product development and operations teams, is seeing more frequent use in the creation of IACS products. To automate the build and delivery processes, DevOps is utilizing a \gls{cicd} pipeline. In such setups, the responsibility for the security assurance of products normally falls within the purview of dedicated security teams, and the security assurance is typically validated by these teams at the final development stage.
Discovering any deviations from the security requirements at this stage can lead to additional expenses as well as causing product releases to be delayed.
To avoid this, a "Shift Left" strategy is typically employed in DevOps to identify and address conformance to security requirements continuously and in the early stages of product development. \Cref{fig:overview} illustrates this practice of incorporating security testing at each level of the software development process, which is known as DevSecOps. Having easy-to-integrate security compliance check tools in early development is key for smoothly and efficiently adopting certification processes. In order to achieve this, appropriate tools and automated testing may assist in reducing requirements for in-depth cybersecurity knowledge and expertise. This will free up development teams to concentrate on product features while supporting their efforts to integrate security by design.

In this work, we present our qualitative analysis of ISA/IEC 62243-4-2 (from here on referred to as \emph{part 4-2})~\cite{iec:62443-4-2} component requirements and list the corresponding types of security tests related to the validation of these requirements. We then present a survey of existing security testing tools, both commercial and open-source, to get a deep understanding of their capabilities and coverage. Finally, we describe our investigation on the tools that can be used in \gls{cicd} for automated testing of part 4-2  security requirements. 

The rest of the paper is organized as follows.
\Cref{sec:background} provides a basic introduction to the standard and discusses the testing methods. 
\Cref{sec:related_work} presents the work related to integration of security tests in DevSecOps.
\Cref{sec:methodology} gives an overview of our methodology for mapping part 4-2 requirements to security tests.
\Cref{sec:tools} presents our analysis of the tooling landscape.
\Cref{sec:findings} describes our findings and the gaps related to tools.
\Cref{sec:conclusion} summarizes this paper and outlines directions for future work.

\section{Background}
\label{sec:background}

The requirements and procedures for addressing the security of \gls{iacs} throughout its lifecycle are specified by the ISA/IEC 62443 set of standards. The standard gives availability of \gls{iacs} more attention in the \gls{cia} triad, since availability is crucial in the \gls{ot} environment as opposed to \gls{it}. The standard provides a set of requirements for significant stakeholder groups involved in the cybersecurity of control systems namely asset owners, suppliers of automation products, integrators building and maintaining \gls{iacs} as well as their components, and service providers assisting with the operation of \gls{iacs}.  
To this end, seven \gls{fr} are defined, which form the basis of the technical requirements of \gls{iacs} and its components. They are namely, (1) Access Control, (2) Use Control, (3) Data Integrity, (4) Data Conﬁdentiality, (5) Restriction of Data Flow, (6) Timely Response to Event, and (7) Resource Availability. 

The standard also defines four security levels (SL1, SL2, SL3, SL4) for evaluating \gls{iacs}, where SL1 is the minimum and SL4 is the maximum level of risk. The security levels are qualitative measures of confidence that the \gls{iacs} can withstand attacks from different classes of threat actors. IEC 62443-4 specifies the secure product development process (part 4-1~\cite{iec:62443-4-1}) as well as  the technical requirements (part 4-2~\cite{iec:62443-4-2}) for \gls{iacs} products. Part 4-2 of the standard breaks down each \gls{fr} into \glspl{cr}. \Gls{cr} may also include a set of \gls{re}. The combination of \gls{cr} and \gls{re} that a component fulﬁlls determines its \gls{sl}.

Fulfilling the \glspl{cr} and \glspl{re} requires understanding of the requirements and mapping them to corresponding security tests that can be carried out through interfaces on the \gls{dut}. The tests thus identified can then be used to validate the device using three different testing methods~\cite{nist:testing}.

\textbf{White-box testing}:
The auditor has complete knowledge of the \gls{dut} as well as access privileges for the device.

\textbf{Black-box testing}:
The internal workings of a system are not known. The tester interacts with the product using externally accessible interfaces  such as graphical user interface, network, or serial communication to validate that the product meets the security requirements. The lack of internal knowledge can make these tests more time-consuming.

\textbf{Gray-box testing}:
This is a compromise between white-box and black-box testing. The tester has some knowledge about the \gls{dut}. This saves time on the reconnaissance phase and focus more on exploiting potential vulnerabilities.

\section{Related work}
\label{sec:related_work}

Various efforts in literature attempted to bridge the gap between DevOps and the security requirements of \gls{iacs} that must obey rigorous requirements from security regulations and standards. However all existing efforts, to the best of our knowledge, fell short of addressing part 4-2 security requirements related tests and mostly focused either on (i) providing schemes to verify secure product development process  for IEC 62443-4-1 or (ii) integrating IEC 62443 security practices into existing engineering processes.

In order to improve the product development process in security regulated environments, diverse papers explore how to automate IEC 62443-4-1 activities through tool support and ensure a DevOps pipeline compliant with the standard~\cite{soares2019, Fabiola, Fabiola2, Bitra, Sadovykh}.
  
A couple of publications focus on the methodology to integrate the security testing and compliance process into the development lifecycle, either at the product development level~\cite{Fockel2019} or at the system integration level~\cite{Maidl2018}. However, our work examines specifically tools that can help this integration of the security testing into the \gls{cicd} pipeline. The following articles are relevant to this goal.

Pfrang \etal~\cite{Pfrang2018} examine different fuzzers for testing IACS. They conclude that none of the investigated tools completely answers industrial systems specificity and propose a new one fulfilling the requirements that they have identified for an ideal \gls{iacs} fuzzer. The proposed fuzzer is integrated into the ISuTest security testing framework for \gls{iacs} that was developed by the authors~\cite{Pfrang2017} to meet these requirements.  %
This work resembles our approach in that it aims at supporting the security testing of \gls{iacs}. However it does not specifically focus on the IEC 62443 security requirements of components (part 4-2) nor on the automation of testing.

Schulz \etal.~\cite{Schulz2018} focus on the evaluation of security testing techniques, looking at available tools from the IT world and their usability for \gls{iacs}. In fact, it discusses exclusively fuzz-testing (for applications, system-call interfaces, system service component, and kernels). Their conclusion is that a security testing suite must be light-weight for evaluation purpose by certification assessors, and maintainable. Fuzzing tools from IT are too complex and necessitate white-box testing.

Leander \etal~\cite{Leander2019} explore how applicable the IEC 62443 standard is to \gls{iiot} systems. \Gls{iiot} systems are basically the combination of \gls{iot} and \gls{iacs}, or in other words this reflects the digital transformation that \gls{iacs} are undergoing. System and component requirements are examined from an IIoT perspective, potential challenges are discussed but no security testing tool is examined. %
Kuli \etal~\cite{Kulik2019} propose an approach to formally verify compliance of cloud-connected SCADA with system security requirements covered by IEC 62443-3-3~\cite{iec:62443-3-3}. A formal modeling language is used to model both the system and the security properties, which can then be formally verified using a model checker. 

Ehrlich\etal~\cite{Ehrlich2019} presents a modelling of all security related functionalities and capabilities with TOSCA specification, which is a language used to define and deploy services in cloud environments. Once the compiler has generated the system model, its compliance with IEC 62443-3-3 can be checked by a security evaluation tool before the generation of Ansible playbooks for automated deployment. %
This work focuses on system requirements and does not report on tools that can support automated testing of component requirements.

Our work, in contrast to the aforementioned works, examines IEC 62243-4-2 requirements and identifies the relevant security tests as well as tools for automated testing of part 4-2 security requirements in \gls{cicd}. Additionally, it also presents the shortcomings in the current tools for achieving full automation and test coverage. 

\section{Methodology}
\label{sec:methodology}

With our methodology we are attempting to capture an accurate and representative subset of today's tooling landscape in order to provide guidance to developers and test engineers for maximizing coverage and automation of conformance with part 4-2 during the \gls{cicd} lifecycle, and also to highlight where tools are currently missing for automatically validating conformance.
To that end we evaluated what configurations, data and information developers and test engineers would have to provide to derive appropriate sets of tools for \gls{cr} and \gls{re} testing at different \glspl{sl}.
As the tooling landscape is vast, we limited ourselves to a representative set of (certified) commercial and open-source tools.
Based on the capabilities and functionalities of these tools, we categorized them according to their use in DevOps lifecycles into the three testing methods black-box, gray-box, and white-box testing.

For each pair of \gls{cr} and tool we estimated the coverage that a tool provides and to which degree testing can be automated for validating conformance with part 4-2.
They infer different levels of knowledge on the inner workings of software as well as hardware and involve different forms of testing.
For black-box testing, we assume no knowledge of the inner workings of a device and we consider both automatic and manual forms of testing.
For gray-box testing, we assume partial knowledge and access to the inner workings of a device as well as the possibility for automated testing.
Finally, we assessed if white-box testing is needed, and in that case we assume that the auditor or tester has full knowledge of a device.
Finally, we considered what kind of output would be required by the different forms of tests and how they would interface with the \gls{cicd} pipeline.
By taking into account all these different factors, we then conducted a qualitative analysis for all \glspl{cr} that combines the type of testing, coverage, degree of automation, as well as set of commercial and open-source tools.
We consider the question of how to actually integrate our qualitative analysis into existing \gls{cicd} pipelines out of scope of this work.

Our assessment consists of two parts.
First, we surveyed existing cybersecurity testing tools (\Cref{sec:tools}), both commercial and open source, to get a deep understanding of their capabilities and limited ourselves to a representative subset.
For tools to be representative we expect that they are being actively maintained and used.
We identified their usage areas (\eg which communication protocols, interface or file types they target), the types of testing they offer (\eg fuzzing, vulnerability scanning), and if they provide an \gls{api} for automation purposes.
Then, we performed a qualitative assessment (\Cref{sec:findings}) to the best of our knowledge for each \gls{cr} in order to determine with respect to the identified tools, which type of testing is possible and what coverage and level of automation is achievable.

Based on our assessment, we have identified and mapped out in our analysis various sets of commercial and open-source tools for validating the conformance of devices with part 4-2.
Using this methodology, we attempt to maximize automation of the validation process while also considering a Shift Left, in particular through the use of open-source tools, to improve time-to-market and accelerate development cycles.
On the one hand, open-source tools have the advantage of being highly customizable which allows for seamless integration into \gls{cicd} pipelines.
Commercial tools, on the other hand, offer specific functionalities which are tailored to the validation process and there are incentives for prolonged maintenance because of usage fees and licenses.
By integrating the best of both worlds into our assessment, we believe to offer optimal guidance in the current tooling landscape.
With our analysis we are providing accurate feedback on the conformance validation while also proposing to further reduce validation time for certification through the use of gray-box and white-box testing methods early in DevOps cycles.

\section{IEC 62443-4-2 Tooling Landscape}

\label{sec:tools}
\vskip.5em

\Gls{isci} is a not-for-profit operational group within ISA.
\Gls{isci} manages the independent ISASecure conformance certification program~\cite{isasecure}, which is based on the IEC security lifecycle defined in IEC 62443.
It defines artifacts that must be provided by certification applicants and tests that need to be run by certification bodies.

\begin{figure}[ht]
    \centering
    \includegraphics[width=.8\linewidth,trim={0.5cm 16.9cm 0.5cm 1cm},clip]{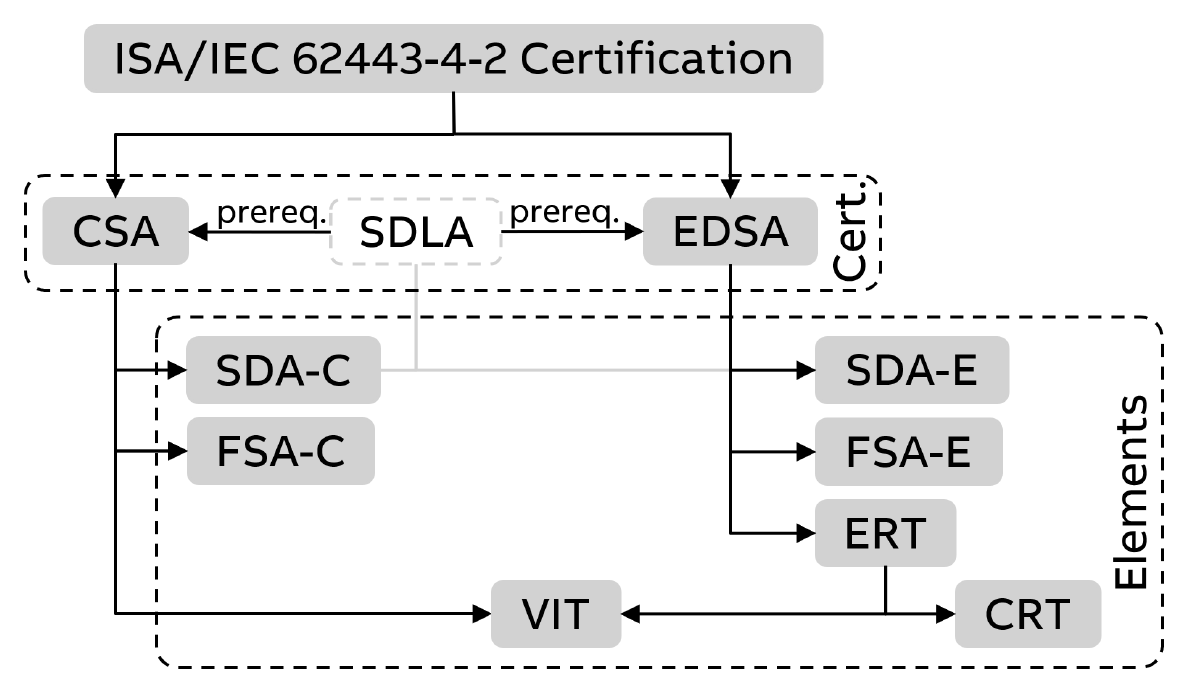}
    \caption{Part 4-2 certifications and their relation to elements.}
    \label{fig:certification}
\end{figure}

The ISASecure certification scheme for product certification of \gls{iacs} compliant with part 4-2 first defines the \gls{sdla}~\cite{isasecure:sdla100} as a mandatory requirement to the \gls{csa}~\cite{isasecure:csa100} and the \gls{edsa}~\cite{isasecure:edsa100}.
\Gls{sdla} itself is ensuring compliance with IEC 62443-4-1.
\Gls{csa} and \gls{edsa} are two versions of part 4-2 certification.
The default one, \gls{csa}, encompasses components such as software applications, host devices, and network devices.
\Gls{edsa} specifically targets embedded devices.
Testing is then subdivided into different elements.
\Gls{sda}~\cite{isasecure:csa312} and \gls{fsa}~\cite{isasecure:csa311} must be performed both for the \gls{csa} (annotated SDA-C and FSA-C) and the \gls{edsa} (SDA-E and FSA-E).
\Gls{vit}~\cite{isasecure:ssa420} is required for \gls{csa} certification.
Additionally, for \gls{edsa}~\cite{isasecure:edsa312} a \gls{crt} must be performed.
The combination of \gls{vit} and \gls{crt} forms the \gls{ert}~\cite{isasecure:edsa310}.
We have visualized dependencies between certifications and their respective testing elements in \cref{fig:certification}.
A non-exhaustive list of companies offering commercial tools for \gls{crt} and \gls{vit} is provided on the ISASecure website.
Testing required for \gls{fsa}, \gls{crt} and \gls{vit} can be partially automated through tools.

The \textbf{\gls{fsa}} is a document mapping \gls{ccsc} and \glspl{cr} of software applications, embedded devices, host devices and network devices to security levels.
For each \gls{ccsc} and \gls{cr} the document describes a requirement as in the part 4-2 standard and which activity is necessary to validate the requirement.
This activity may or may not require the auditor to conduct an independent test.
However, these tests depend on the interfaces the \gls{dut} is exposing, which involves manual testing, limiting the use of tools for automation, and may require very specific tools.

The robustness of devices and systems against network attacks is ensured with \textbf{\gls{crt}}.
During \gls{crt}, \gls{dut} are being evaluated for common programming errors and \gls{dos} vulnerabilities.
Identification and mitigation against malicious code is not within the scope of these tests but is covered by \gls{sdla}.
Product vendors must carry out the testing themselves as the certification body will not perform any \gls{crt} during the certification audition.

\textbf{\Gls{vit}} consists in scanning the \gls{dut} to check for known vulnerabilities from the \gls{nvd}~\cite{nist:nvd}.
This database is continuously updated as new vulnerabilities are identified and verified.
It organizes vulnerabilities into globally accepted \gls{cwe} categories~\cite{mitre:cwe}, which are types of software and hardware weaknesses.
A subset of issues identified under the \gls{cvss} threshold must either be corrected or reported, depending on the \gls{sl} the tool is certified for.

\subsection{Commercial Security Tools}
\label{subsec:commercial}

The following paragraphs summarize a list of officially recognized tools mentioned in \gls{isci} press releases~\cite{isasecure:press}.

\nracheron~\textbf{Acheron}~\cite{cncert:acheron} by CNCERT/CC and Beijing Xinlian Kehui Technology is a \gls{crt} tool that can be used for injecting errors, manipulating packets, altering contexts, and stress testing for more than 50 Internet and industrial Internet protocols.
Furthermore, the tool is capable of monitoring \glspl{dut} using various protocols or alternate methods.
Acheron offers automated testing as well as recording traces when \glspl{dut} show abnormal behavior and can remotely reset \glspl{dut} if they fail.
Custom protocols can be used by the tool, which relies on passive traffic analysis, machine learning, and data analysis in order to automatically derive protocol specifications.

\nrachilles~\textbf{Achilles}~\cite{gedigital:achilles} from GE Digital is  a test platform for monitoring networks and operational parameters as well as discovering vulnerabilities and reproducing faults.
Tests can be automated within the framework by using a dedicated device that is sending test traffic to \gls{dut}.
These tests evaluate protocol boundary conditions by sending invalid, malformed or unexpected packets.
Load testing can be done for various protocols in order to determine the threshold for \gls{dut} under \gls{dos} attack scenarios.
A full set of tests can be run by the test platform in two hours.

\nrbestorm~\textbf{beSTORM}~\cite{beyondsecurity:bestorm} from Beyond Security is a black-box fuzzer that is providing support for over 250 prebuilt modules and protocols.
The tool has the capability to learn and tests augmented, proprietary and custom protocols.
It is possible to automatically monitor and scan systems for vulnerabilities either on-site or with a cloud-based solution.
Tests can also be conducted offline and results can be exported for further analysis.
The tool has been designed for reducing the number of false positive results.

\nrdefensics~\textbf{Defensics}~\cite{synopsys:defensics,synopsys:defensicsISASecure} is a black-box protocol fuzzer by Synopsys that uses fuzz testing in a negative testing approach to identify issues in the \gls{sut}.
The tool can distinguish between interface, protocol and file format types for running targeted test cases against systems.
With more than 300 prebuilt generational tests suites, Defensics is capable of covering a wide range of protocols and allows customization of any test suite.
A \gls{sdk} written in Java can be used for implementing support of proprietary or custom input types and extend the capabilities of the tool.
Issues identified by the tool can be recreated to assist in identification of the origin of the issue as well as later for validation of the fixed issue.

\nrnessus~\textbf{Nessus}~\cite{tenable:nessus} is a long-running commercial vulnerability scanner by Tenable and being used by \gls{isci} in order to run the \gls{vit}.
The tool provides tests covering many \glspl{cwe}, support for dynamically compiled plug-ins and can even be deployed on a Raspberry Pi~\cite{tenable:rpi}.
Data from previous scans is being collected and used for assessing new vulnerabilities without having to run an actual scan.
Scan reports can be exported in various formats for sharing and further analysis.

\nrraven~\textbf{Raven ES}~\cite{fujisoft:raven} from FUJISOFT is a product security testing suite originally developed by Hitachi Systems and then FFRI Security.
The tool uses fuzz testing and monitors the behavior of the \gls{dut} by sending regular data to check if the device is still operational.
Any crashes or unexpected behavior of the \gls{dut} are considered as identified vulnerabilities.
Support for many Internet and industrial Internet protocols is provided by the tool, which makes it usable by non-security experts.
One inconvenience of the tool is the Japanese user interface that hinders the ease of use.

\nrvhunter~\textbf{VHunter IVM}~\cite{winicssec:vhunterivm} by Beijing Winicssec Technologies is an industrial control vulnerability mining platform used as \gls{crt} tool that supports \gls{ert} evaluation through black-box testing.
The platform consists of a device that is connected to the network of the \gls{dut} in order to automatically detect vulnerabilities, run stress tests, and reset the \gls{dut} by interfacing with its power supply if necessary.
An extensive list of protocol fuzz tests are covered by the platform for a large number of Internet and industrial Internet protocols.

\subsection{Open-Source Security Alternatives}
\label{subsec:oss}

Many alternative solutions to commercial tools have surfaced in the open-source community which are not necessarily specific to the standard.
The following paragraphs provide a list of cybersecurity tools that appear the most relevant to us for testing compliance with part 4-2 in \gls{cicd} pipelines.
This list is representing a subset of tools we consider relevant for the scope of our analysis and is by no means complete.

\nrafl~\textbf{AFL++}~\cite{aflplusplus} is an open-source fork of the original, but no longer maintained, \gls{afl} fuzzer developed by Michał Zalewski.
\Gls{afl} has had a strong influence on fuzzing as a research domain and many fuzzers did emerge as forks of AFL~\cite{nichols:neural,pham:aflnet,wang:superion,zheng:firmafl}.
As a gray-box fuzzer, AFL++ adds instrumentation to the binary that is to be fuzzed and explores it using genetic algorithms.
The input is provided by the user and fed to the binary over standard input, which restricts the interface by which the family of \gls{afl} fuzzers interact with binaries.
However, there are ways to lift these restrictions and interface with binaries over sockets, such as by using for example Preeny~\cite{preeny}, which employs preload patching to redirect I/O streams by modifying I/O system calls.

\nrburpsuite~\textbf{Burp Suite}~\cite{portswigger:burpsuite} is a security testing tool by PortSwigger that is available as a free community edition or as multiple tiered commercial editions with additional features.
The tool acts as a proxy and intercepts all HTTP(S) and WebSocket traffic that is being exchanged between a browser and a \gls{dut}.
This allows comparing, inspecting and manipulating messages at various stages during the connection.
Additional features such as Repeater, Intruder, and many more provide functionalities that go beyond the basic proxy functionality and can be used for testing and identifying very specific vulnerabilities.
In particular Intruder enables automation of operations such that targeted attacks can be run on the \gls{dut}.

\nrjmeter~\textbf{JMeter}~\cite{apache:jmeter} is an open-source project under the Apache Software Foundation.
The tool can be used for performance and load testing for various services and protocols.
Tests can be run either from a \gls{cli} or from an \gls{ide}, whereas the \gls{ide} provides additional features such as recording, building and debugging.
Response formats such as HTTP, JSON, XML and other text formats are supported and results can be cached or stored for offline analysis as well as being replayed.
Plugins and scripts for samplers and visualisation allow extending the tool and also allow for integration into the \gls{cicd} pipeline.

\nrnmap~\textbf{nmap}~\cite{nmap} is an open-source tool for detecting nodes and services in networks and for security auditing.
Raw IP packets are being used by the tool in novel ways as identification mechanism for nodes, services, \glspl{os}, packet filters and firewalls to name a few.
Users are presented with a list of ports for each scanned node indicating which protocol is used and what service and version is running.
A scripting engine allows automating operations written in the Lua~\cite{lua} programming language.

\nropenvas~\textbf{OpenVAS}~\cite{greenbone:openvas} is an open-source  \gls{vit} framework from Greenbone that allows performing authenticated and unauthenticated testing.
The framework provides support for many Internet and industrial Internet protocols as well as support for implementing custom vulnerability tests.
A vulnerability manager daemon, security assistant daemon (web interface) and scanner executable form the framework which is also known as \gls{gvm}.
The Security Assistant can be controlled over interfaces or external tools using the Greenbone Management Protocol.
Commercial offerings include enterprise appliances that contain the \gls{gvm} framework as well as an \gls{os} with additional functionalities.
It is possible to subscribe the framework to either a community or an enterprise feed which are continuously updated to receive tests for detecting vulnerabilities.
A cloud service as software-as-a-service solution is also available.

\nrscapy~\textbf{Scapy}~\cite{scapy} is an open-source Python library and tool for manipulating packets.
It provides support for many protocols and relies on pcap~\cite{tcpdump} for storing and reading packets in files.
Requests can be matched to responses and it is possible to inject invalid frames and packages.

\nrselenium~\textbf{Selenium}~\cite{selenium} is an open-source umbrella project for a set of tools to automate web tests.
WebDriver provides the language bindings for various programming languages and controlling component for supported browsers, that can run web tests either locally or remotely.
The \gls{ide} tool is a browser extension that allows recording web tests for supported browsers, which are stored as scripts written in the Selenese domain-specific language.
It is possible to operate the \gls{ide} using the command line interface or to extend it through plugins.
Tests can be distributed and executed in parallel using the Grid tool. 

\nrsymcc~\textbf{SymCC}~\cite{poeplau2020symcc} is an open-source compiler that leverages concolic execution~\cite{sen:cute} for finding all possible execution paths.
The LLVM-based compiler instruments code at compile-time in order to interface at run-time with a satisfiability modulo theories~\cite{deMoura:z3} solver and speed up symbolic execution over alternative tools that make use of an interpreter approach.
Combining SymCC with AFL allows to perform hybrid fuzzing.

\section{Analysis}
\label{sec:findings} %

The goal of our qualitative analysis is to provide developers and test engineers with a representative summary of the current tooling landscape for conformance validation automation of part 4-2 (taking into account the subsets of tools covered in this study).
We address in particular two main questions: (1) which commercial and open-source tools would be necessary for covering the required tests and (2) to what extent can these tests be integrated into the \gls{cicd} pipeline.
Moreover, our analysis highlights the areas where support for tooling can be improved or needs to be expanded in order to reach better coverage and automation of the conformance validation.

\Cref{tab:ci-testing} shows the result of our qualitative analysis describing the current state of automating conformance validation within \gls{cicd} pipelines with respect to part 4-2.
For each \gls{cr}, \gls{re}, and \gls{sl} triple, we determined the supported types of testing as well as tuples of the form $\mathcal{T} = (\mathcal{A},\mathcal{C},\mathcal{O})$ for each of the three testing methods, namely white-box, gray-box, and black-box testing.
The types of testing are represented in a state register format in alphabetic order:
\begin{itemize}
    \item authentication (\texttt{A}): the tool uses some interface or protocol to authenticate itself to the component,
    \item fuzzing (\texttt{F}): input data boundary testing,
    \item fault injection (\texttt{I}): injection of incorrect data with respect to a protocol specification,
    \item monitoring (\texttt{M}): the tool detects that the component has crashed or gone into a faulty state,
    \item performance (\texttt{P}): the tool measures the time to response and/or detects \gls{dos},
    \item vulnerability (\texttt{V}) testing: checking behavior against known Common Vulnerabilities and Exposures (CVEs).
\end{itemize}
We represent the testing type in the state register by a dash (\texttt{-}), whenever that type of testing is not covered by the sets of tools in the tuples.

The overall coverage is indicated by the first element in the tuple $\mathcal{A}\in\{\mathrm{\textcovered},\mathrm{\textcovpartially},\mathrm{\textuncovered}\}$, which is the union of the coverage we assessed during the first part with our methodology for each tool in the sets of commercial and open-source tools.
The coverage is subdivided into mostly covered and automated (\textcovered), partially covered and automated (\textcovpartially), and not covered and manual action required (\textuncovered).
Commercial ($\mathcal{C}\subseteq$\{\nracheron,$\ldots$,\nrvhunter\}) and \gls{oss} ($\mathcal{O}\subseteq$\{\nrafl,$\ldots$,\nrsymcc\}) tools (see \cref{subsec:commercial,subsec:oss} for descriptions of tools) could be used for implementing mostly or partially automated tests.

For example, in the case of \gls{fr} 3 \emph{System integrity} for \gls{cr} 5 \emph{Input validation}, it is required that the syntax, length and content of input for all external interfaces is validated.
The tooling landscape covers most types of testing with tuples for all stages of the DevOps lifecycle: (\textcovered, $\varnothing$, \{\nrsymcc\}), (\textcovpartially, $\varnothing$, \{\nrafl\}), and (\textcovered,\{\nrbestorm,\nrdefensics\},$\varnothing$).
Authentication, fuzz, fault injection, monitoring, and performance testing are the types of test that can be executed given the sets of commercial and open-source tools in the three tuples.
For white-box testing mostly covered and automated (\textcovered) for the \gls{oss} tool \nrsymcc, even though no suitable commercial tool ($\varnothing$) was identified.
Through concolic execution all interface execution paths can be explored and evaluated.
Gray-box testing is only partially covered and automated (\textcovpartially) for the \gls{oss} tool \nrafl,
mainly because additional instrumentation is necessary in order to fuzz the interface using AFL++.
Lastly, black-box testing is mostly covered and automated (\textcovered) for the set of commercial tools \{\nrbestorm,\nrdefensics,\nrraven,\nrvhunter\}, which make use of different fuzzing techniques.

For \gls{fr} 2 \gls{cr} 13, \gls{fr} 3 \glspl{cr} 11 to 14, \gls{fr} 4 \gls{cr} 2, as well as \gls{fr} 7 \glspl{cr} 3, 4, 6, 7, and 8 we were not able to identify tools that could be used for validating conformity. This is represented in the table by an empty line. Other types of testing or investigation would then be needed. For example, to check compliance with \gls{fr} 7 \gls{cr} 7 \emph{Least functionality}, one has to manually check that it is possible to deactivate unnecessary functions, ports, protocols and/or services.

A line with the "Not applicable (N/A)" indicates that part 4-2 does not provide a component level requirement associated with the corresponding system requirement from part 3-3.
\begin{table}[!t]
    \centering
    \caption{Qualitative assessment of IEC 62443 requirement testing.\label{tab:ci-testing}}
    \setlength{\aboverulesep}{0pt}
    \setlength{\belowrulesep}{0pt}
    \setlength{\tabcolsep}{3pt}
    \rowcolors{1}{gray!10}{gray!0}
    {\scriptsize%
    \begin{tabular}{>{\kern-\tabcolsep}l|c|c|c|c<{\kern-\tabcolsep}}
        \toprule\rowcolor{gray!25}
        \multicolumn{1}{c|}{\textbf{Req.}} & \textbf{Type} & \textbf{White-box} & \textbf{Gray-box} & \textbf{Black-box} \\
        \midrule
        \textbf{CCSC} & & & & \\
        1A & \texttt{AFIMPV} & & (\covpartially,$\varnothing$,\{\nrburpsuite,\nrjmeter,\nrnmap,\nrscapy,\nrselenium\}) & (\covpartially,\{\nracheron,\nrachilles,\nrbestorm,\nrdefensics,\nrnessus,\nrraven,\nrvhunter\},\{\nropenvas\}) \\
        1D & \texttt{AFIMPV} & & (\covpartially,$\varnothing$,\{\nrburpsuite,\nrjmeter,\nrnmap,\nrselenium\}) & (\covpartially,\{\nracheron,\nrachilles,\nrbestorm,\nrdefensics,\nrnessus,\nrvhunter\},\{\nropenvas\}) \\
        \midrule
        \textbf{FR 1} & & & & \\
        CR 1 & \texttt{AFIMPV} & & (\covpartially,$\varnothing$,\{\nrburpsuite,\nrjmeter,\nrnmap,\nrselenium\}) & (\covpartially,\{\nracheron,\nrachilles,\nrbestorm,\nrdefensics,\nrnessus\},\{\nropenvas\}) \\
        CR 2 & \texttt{AFIMPV} & & (\covpartially,$\varnothing$,\{\nrburpsuite,\nrjmeter,\nrnmap,\nrselenium\}) & (\covpartially,\{\nracheron,\nrachilles,\nrbestorm,\nrdefensics,\nrnessus,\nrvhunter\},\{\nropenvas\}) \\
        CR 3 & \texttt{AFIMPV} & & (\covpartially,$\varnothing$,\{\nrburpsuite,\nrjmeter,\nrnmap,\nrscapy,\nrselenium\}) & (\covpartially,\{\nracheron,\nrachilles,\nrbestorm,\nrdefensics,\nrnessus\},\{\nropenvas\}) \\
        CR 4 & \texttt{AFIMPV} & & (\covpartially,$\varnothing$,\{\nrjmeter,\nrscapy,\nrselenium\}) & (\covpartially,\{\nracheron,\nrachilles,\nrdefensics,\nrnessus\},\{\nropenvas\}) \\
        CR 5 & \texttt{AFIMPV} & & (\covpartially,$\varnothing$,\{\nrburpsuite,\nrjmeter,\nrselenium\}) & (\covpartially,\{\nracheron,\nrachilles,\nrbestorm,\nrdefensics,\nrnessus\},\{\nropenvas\}) \\
        CR 6 & \texttt{AFIMPV} & & (\covpartially,$\varnothing$,\{\nrscapy\}) & (\covpartially,\{\nrbestorm,\nrdefensics,\nrnessus\},$\varnothing$) \\
        CR 7 & \texttt{AFIMPV} & & (\covpartially,$\varnothing$,\{\nrburpsuite,\nrselenium\}) & (\covpartially,\{\nracheron,\nrachilles,\nrbestorm,\nrdefensics,\nrnessus\},$\varnothing$) \\
        CR 8 & \texttt{AFIMPV} & & (\covpartially,$\varnothing$,\{\nrburpsuite,\nrjmeter,\nrscapy,\nrselenium\}) & (\covered,\{\nracheron,\nrachilles,\nrdefensics,\nrnessus\},$\varnothing$) \\
        CR 9 & \texttt{AFIMPV} & & (\covpartially,$\varnothing$,\{\nrburpsuite,\nrscapy\}) & (\covered,\{\nracheron,\nrachilles,\nrdefensics,\nrnessus\},\{\nropenvas\}) \\
        CR 10 & \texttt{AFIMPV} & & (\covpartially,$\varnothing$,\{\nrburpsuite,\nrselenium\}) & (\covpartially,\{\nracheron,\nrachilles,\nrbestorm,\nrdefensics\},$\varnothing$) \\
        CR 11 & \texttt{AFIMPV} & & (\covpartially,$\varnothing$,\{\nrburpsuite,\nrjmeter,\nrnmap,\nrscapy,\nrselenium\}) & (\covered,\{\nracheron,\nrachilles,\nrbestorm,\nrdefensics,\nrraven\},$\varnothing$) \\
        CR 12 & \texttt{AFIMPV} & & (\uncovered,$\varnothing$,\{\nrburpsuite,\nrselenium\}) & (\covpartially,\{\nracheron,\nrachilles,\nrbestorm,\nrdefensics\},$\varnothing$) \\
        CR 13 & \texttt{AF-MPV} & & (\covpartially,$\varnothing$,\{\nrjmeter\}) & (\covpartially,\{\nrbestorm,\nrdefensics,\nrnessus\},$\varnothing$) \\
        CR 14 & \texttt{AFIMP-} & & (\covpartially,$\varnothing$,\{\nrnmap,\nrscapy\}) & (\covpartially,\{\nrdefensics\},$\varnothing$) \\
        \midrule
        \textbf{FR 2} & & & & \\
        CR 1 & \texttt{AFIMPV} & & (\covpartially,$\varnothing$,\{\nrburpsuite,\nrjmeter,\nrselenium,\nrscapy\}) & (\covered,\{\nracheron,\nrachilles,\nrdefensics,\nrnessus\},$\varnothing$) \\
        CR 2 & \texttt{AFIMPV} & & (\covpartially,$\varnothing$,\{\nrscapy\}) & (\covered,\{\nrbestorm,\nrdefensics,\nrnessus\},$\varnothing$) \\
        CR 3 & N/A & N/A & N/A & N/A \\
        CR 4 & \texttt{AFIMPV} & & (\covpartially,$\varnothing$,\{\nrjmeter,\nrselenium\}) & (\covpartially,\{\nracheron,\nrbestorm,\nrdefensics,\nrnessus\},$\varnothing$) \\
        CR 5 & \texttt{AFIMPV} & & (\covpartially,$\varnothing$,\{\nrburpsuite,\nrjmeter,\nrnmap,\nrselenium\}) & (\covered,\{\nracheron,\nrachilles,\nrbestorm,\nrdefensics,\nrnessus,\nrvhunter\},$\varnothing$) \\
        CR 6 & \texttt{AFIMPV} & & (\covpartially,$\varnothing$,\{\nrjmeter\}) & (
        (\covered,\{\nracheron,\nrachilles,\nrbestorm,\nrdefensics,\nrnessus\},$\varnothing$) \\
        CR 7 & \texttt{AFIMPV} & & (\covpartially,$\varnothing$,\{\nrburpsuite,\nrjmeter,\nrnmap,\nrselenium\}) & (\covered,\{\nracheron,\nrachilles,\nrbestorm,\nrdefensics,\nrnessus,\nrvhunter\},$\varnothing$) \\
        CR 8 & \texttt{AFIMPV} & & & (\covered,\{\nracheron,\nrachilles,\nrbestorm,\nrdefensics\},$\varnothing$) \\
        CR 9 & \texttt{AFIMPV} & & & (\covered,\{\nracheron,\nrachilles,\nrbestorm,\nrdefensics\},$\varnothing$) \\
        CR 10 & \texttt{AFIMPV} & & & (\covered,\{\nracheron,\nrachilles,\nrbestorm,\nrdefensics\},$\varnothing$) \\
        CR 11 & \texttt{AFIMPV} & & (\uncovered,$\varnothing$,\{\nrjmeter\}) & (\covered,\{\nracheron,\nrachilles,\nrdefensics\},$\varnothing$) \\
        CR 12 & \texttt{AFIMPV} & & & (\covpartially,\{\nracheron,\nrachilles,\nrdefensics\},$\varnothing$) \\
        CR 13 & \texttt{-}\texttt{-}\texttt{-}\texttt{-}\texttt{-}\texttt{-} & & & \\
        \midrule
        \textbf{FR 3} & & & & \\
        CR 1 & \texttt{AFIMPV} & & (\covpartially,$\varnothing$,\{\nrburpsuite,\nrjmeter,\nrscapy,\nrselenium\}) & (\covered,\{\nracheron,\nrachilles,\nrbestorm,\nrdefensics\},$\varnothing$) \\
        CR 2 & \texttt{AFIMPV} & & (\covpartially,$\varnothing$,\{\nrselenium\}) & (\covpartially,\{\nracheron,\nrachilles,\nrdefensics,\nrnessus\},\{\nropenvas\}) \\
        CR 3 & \texttt{AFIMPV} & & (\covpartially,$\varnothing$,\{\nrselenium\}) & (\covered,\{\nracheron,\nrachilles,\nrbestorm,\nrdefensics,\nrraven\},$\varnothing$) \\
        CR 4 & \texttt{AFIMPV} & & & (\covpartially,\{\nracheron,\nrachilles,\nrdefensics\},$\varnothing$) \\
        CR 5 & \texttt{AFIMP-} & (\covered,$\varnothing$,\{\nrsymcc\}) & (\covpartially,$\varnothing$,\{\nrafl\}) & (\covered,\{\nrbestorm,\nrdefensics,\nrraven,\nrvhunter\},$\varnothing$) \\
        CR 6 & \texttt{AFIMPV} & & & (\covpartially,\{\nracheron,\nrachilles,\nrdefensics,\nrnessus\},\{\nropenvas\}) \\
        CR 7 & \texttt{AFIMPV} & & & (\covpartially,\{\nracheron,\nrachilles,\nrdefensics,\nrnessus\},\{\nropenvas\}) \\
        CR 8 & \texttt{AFIMPV} & & (\covpartially,$\varnothing$,\{\nrburpsuite,\nrjmeter,\nrscapy,\nrselenium\}) & (\covered,\{\nracheron,\nrachilles,\nrbestorm,\nrdefensics\},$\varnothing$) \\
        CR 9 & \texttt{AFIMPV} & & & (\covpartially,\{\nracheron,\nrachilles,\nrdefensics\},$\varnothing$) \\
        CR 10 & \texttt{A-}\texttt{-MP-} & & (\uncovered,$\varnothing$,\{\nrselenium\}) & \\
        CR 11 & \texttt{-}\texttt{-}\texttt{-}\texttt{-}\texttt{-}\texttt{-} & & & \\
        CR 12 & \texttt{-}\texttt{-}\texttt{-}\texttt{-}\texttt{-}\texttt{-} & & & \\
        CR 13 & \texttt{-}\texttt{-}\texttt{-}\texttt{-}\texttt{-}\texttt{-} & & & \\
        CR 14 & \texttt{-}\texttt{-}\texttt{-}\texttt{-}\texttt{-}\texttt{-} & & & \\
        \midrule
        \textbf{FR 4} & & & & \\
        CR 1 & \texttt{AFIMPV} & & (\covpartially,$\varnothing$,\{\nrburpsuite,\nrjmeter,\nrscapy,\nrselenium\}) & (\covpartially,\{\nracheron,\nrachilles,\nrbestorm,\nrdefensics\},$\varnothing$) \\
        CR 2 & \texttt{-}\texttt{-}\texttt{-}\texttt{-}\texttt{-}\texttt{-} & & & \\
        CR 3 & \texttt{AFIMPV} & & (\covpartially,$\varnothing$,\{\nrburpsuite\}) &  (\covered,\{\nracheron,\nrachilles,\nrbestorm,\nrdefensics,\nrnessus\},\{\nropenvas\}) \\
        \midrule
        \textbf{FR 5} & & & & \\
        CR 1 & \texttt{AFIM}\texttt{-}\texttt{-} & & (\covpartially,$\varnothing$,\{\nrnmap\}) & (\covpartially,\{\nrbestorm,\nrdefensics\},$\varnothing$) \\
        CR 2 & \texttt{AFIM}\texttt{-}\texttt{-} & & (\covpartially,$\varnothing$,\{\nrnmap\}) & (\covpartially,\{\nrbestorm,\nrdefensics\},$\varnothing$) \\
        CR 3 & \texttt{AFIMP-} & & (\covpartially,$\varnothing$,\{\nrnmap,\nrscapy\}) & (\covpartially,\{\nrbestorm,\nrdefensics\},$\varnothing$) \\
        CR 4 & N/A & N/A & N/A & N/A \\
        \midrule
        \textbf{FR 6} & & & & \\
        CR 1 & \texttt{AFIMPV} & & & (\covpartially,\{\nracheron,\nrachilles,\nrbestorm\nrdefensics\},$\varnothing$) \\
        CR 2 & \texttt{A}\texttt{-}\texttt{-}\texttt{-}\texttt{-}\texttt{V} & & & (\covpartially,\{\nrnessus\},$\varnothing$) \\
        \midrule
        \textbf{FR 7} & & & & \\
        CR 1 & \texttt{AFIMPV} & & (\covpartially,$\varnothing$,\{\nrjmeter,\nrnmap\}) & (\covpartially,\{\nrnessus,\nrraven\},$\varnothing$) \\
        CR 2 & \texttt{AFIMP-} & & & (\covpartially,\{\nrraven,\nrvhunter\},$\varnothing$) \\
        CR 3 & \texttt{-}\texttt{-}\texttt{-}\texttt{-}\texttt{-}\texttt{-} & & & \\
        CR 4 & \texttt{-}\texttt{-}\texttt{-}\texttt{-}\texttt{-}\texttt{-} & & & \\
        CR 5 & N/A & N/A & N/A & N/A \\
        CR 6 & \texttt{-}\texttt{-}\texttt{-}\texttt{-}\texttt{-}\texttt{-} & & & \\
        CR 7 & \texttt{-}\texttt{-}\texttt{-}\texttt{-}\texttt{-}\texttt{-} & & & \\
        CR 8 & \texttt{-}\texttt{-}\texttt{-}\texttt{-}\texttt{-}\texttt{-} & & & \\
        \bottomrule
    \end{tabular}%
    \\
    \begin{tcolorbox}[colback=LegendBack,colframe=LegendBorder,arc=6pt,left=2pt,right=2pt,top=1pt,bottom=1pt,boxsep=0pt,width=.85\linewidth,before skip=.29em,after skip=0em]
    \rowcolors{1}{LegendBack}{LegendBack}
    \begin{tabular}{llclcl}
    \textbf{Legend:} & Types & \texttt{A} & Authentication & \texttt{M} & Monitoring \\
     & & \texttt{F} & Fuzzing & \texttt{P} & Performance \\
     & & \texttt{I} & Fault injection & \texttt{V} & Vulnerability testing \\
     & Tuples & $\mathcal{T}$ & \multicolumn{3}{l}{(\textit{overall coverage}, \textit{comm. tools}, \textit{OSS tools})} \\
     & Coverage & $\mathcal{A}$ & \multicolumn{3}{l}{\covered~mostly / \covpartially~partially / \uncovered~not covered} \\
     & Tools & $\mathcal{C}$ & \textit{commercial} & \multicolumn{2}{l}{$\{$\nracheron,$\ldots$,\nrvhunter$\}$} \\
     & & $\mathcal{O}$ & \textit{OSS} & \multicolumn{2}{l}{$\{$\nrafl,$\ldots$,\nrsymcc$\}$} \\
    \end{tabular}
    \end{tcolorbox}
    }%
\end{table}

\subsection{The OT Gap of Security Tools}

Tools mentioned in \cref{sec:tools} have interesting features and can assist in part 4-2 security testing to make it part of \gls{cicd} pipelines. Though, they may come with some shortcomings to be aware of. Furthermore, it should be noticed that some \glspl{cr} cannot have or do not require tooling support. This is understandable given that the evaluation of these \glspl{cr} might require different kinds of analysis. For example, \gls{fr} 3 \gls{cr} 11 \emph{Physical tamper resistance and detection} which requires manual inspection of the product.

Most open-source tools (for security testing) mentioned in \cref{subsec:oss} come from the IT world and usually do not support protocols, programming languages, \etc that are used in the OT world or which are specific to industrial components. Consequently, there may be aspects of IACS products that could possibly not be covered by these tools. These tools should, however, enable code modification and plugin development albeit with considerable effort. Additionally, because open-source security testing tools are not certified, it is challenging to evaluate their quality. Therefore, the choice between commercial and open-source tools must take into account the final cost and quality.

One challenge with \gls{iacs} is that their interfaces can sometimes not be automated, which is a pre-requirement for enabling automated testing.
Depending on \gls{hmi} support, \gls{cicd} testing may be possible or difficult, and in cases where \gls{cicd} testing is impossible, human testing would be required for asserting conformance.

\subsection{Incomplete Coverage of OSS Alternatives}

Certain compliance checks may not be available in open-source tools.
For example, Nessus is the only \gls{vit} tool certified to conform with IEC 62443.
While Nessus and its \gls{oss} alternative OpenVAS do have many checks in common, neither of them fully covers the set of tests of the other.
Therefore there is no certainty that the entire pipeline needed for automatic testing can be covered by only \gls{oss} tools. 
To reach similar scope and capabilities of commercial tools, it is necessary to combine several of their \gls{oss} counterparts and possibly extend them as mentioned in the previous paragraph.

\subsection{OSS Tools' Short Lifetime}

The \gls{oss} alternatives specific to ISASecure that we have identified show stability issues and are not necessarily well maintained. 
Among these tools are for example hping3~\cite{sanfilippo:hping} or HULK~\cite{shteiman:hulk}, which are no longer actively maintained, but for which features were integrated into other, actively maintained tools or forks~\cite{aamirkhan:ihulk}. It is interesting to note that hping3, a TCP/IP packet forging tool with scripting functionalities, is still being used and recommended by the security community~\cite{okta:hping3, securium:hping3}.
In general we observed that \gls{oss} tools are much more short living than the lifetime of \gls{iacs}.
Such tools may still be useful, though, but to support \gls{cicd}-integrated testing would require the development team to tune and maintain these tools themselves. This is contradictory with our aim of making the integration of security testing into the \gls{cicd} pipeline as easy and as automated as possible.

\section{Conclusion and Future Work}
\label{sec:conclusion}

This paper presented an extensive qualitative analysis of the IEC 62443-4-2 standard requirements in order to better understand how to design an automated conformance validation process within CI/CD development pipelines of industrial automation and control systems. As part of the analysis, the paper demonstrated the coverage of commercial as well as \gls{oss} testing tools that are currently available for validating 
 part 4-2 security component requirements. The paper identified the relevant tools as well as the stages in the CI/CD pipeline where those tools can be used. Our analysis showcased the following interesting observations:
\begin{itemize}
    \item commercial tools used for part 4-2 certification are mainly built using black-box testing methods
    \item open-source tools provide gray-box testing methods, however, they cannot be directly applied to OT products
    \item tools for part 4-2 white-box testing are very scarce
    \item validation of some part 4-2 requirements needs inherent manual intervention or  development of custom-tailored tools and tests.  
\end{itemize}

An interesting open future research direction is exploring the possibility of building a framework that unifies and orchestrates the execution of all those different tools needed for part 4-2 verification. This framework would seamlessly integrate such tools within CI/CD pipelines and consolidate the reporting of these tools into a unified security assessment report. Such reports can be used by security experts as well as for conformance certification at later stages.  
 
{\small
\bibliographystyle{IEEEtran}
\bibliography{IEEEabrv,references}
}

\end{document}